\documentclass{mem}
\usepackage{natbib}\usepackage{txfonts}\usepackage{balance}
\usepackage{graphicx}
\usepackage[a4paper,breaklinks,dvipdfm]{hyperref}
\idline{84}{1}
\usepackage{txfonts} 
\def\XMM{\textit{XMM--Newton}}

\begin{document}

\title{Blackbody excess in persistent Be pulsars}

\author{N. \,La Palombara\inst{1}, S. \,Mereghetti\inst{1}, L. \,Sidoli\inst{1}, A. \,Tiengo\inst{1,2} \and P. \,Esposito\inst{1}}

\institute{INAF - IASF Milano, via Bassini 15, I--20133 Milano, Italy
\and
IUSS, Piazza della Vittoria 15, I--27100 Pavia, Italy}

\authorrunning{La Palombara}

\titlerunning{Blackbody excess in persistent Be/NS binary pulsars}

\abstract{We report on the main results obtained thanks to an observation campaign, performed with \XMM, of four persistent, low--luminosity ($L_{\rm X} \sim 10^{34}$ erg s$^{-1}$) and long--period ($P > 200$ s) Be accreting pulsars. We found that all sources considered here are characterized by a spectral excess that can be described with a blackbody component of high temperature ($kT_{\rm BB} > 1$ keV) and small area ($R_{\rm BB} < 0.5$ km). We show that: 1) this feature is a common property of several low--luminosity X--ray binaries; 2) for most sources the blackbody parameters (radius and temperature) are within a narrow range of values; 3) it can be interpreted as emission from the NS polar caps.
}
\maketitle{}

We have analyzed the \XMM\ observations of the four \textit{persistent} Be pulsars originally identified by \citet{Reig&Roche99}, i.e. \mbox{RX J0146.9+6121} \citep{LaPalombara&Mereghetti06}, \mbox{4U 0352+309} \citep{LaPalombara&Mereghetti07}, \mbox{RX J1037.5-5647} \citep{LaPalombara+09}, and \mbox{RX J0440.9+4431} \citep{LaPalombara+12}; their main parameters are reported in Table~\ref{parameters}. These sources have persistently low luminosity ($L_{\rm X} \sim 10^{34-35}$ erg s$^{-1}$) and long pulse period ($P > 200$ s), two properties which suggest that the neutron star (NS) orbits the Be star in a wide and nearly circular orbit, continuously accreting material from the low--density outer regions of the circumstellar envelope.

\begin{figure*}[]
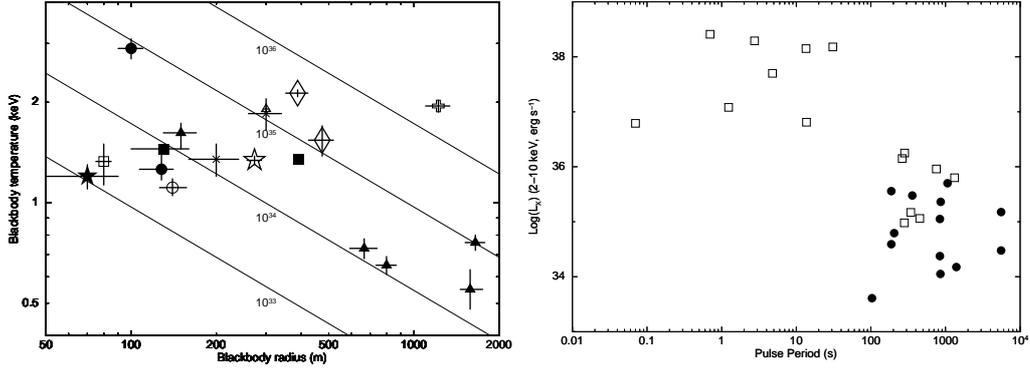

\resizebox{16pc}{!}{\includegraphics[angle=-90]{LaPalombaraN_f1.ps}}
\resizebox{16pc}{!}{\includegraphics[angle=-90]{LaPalombaraN_f2.ps}}
\caption{
\footnotesize
\textit{Left}: Best--fit values for $R_{\rm BB}$ and $kT_{\rm BB}$ of the low--luminosity high--mass X--ray binaries with a \textit{hot--BB} spectral component (different symbols refer to different sources); the continuous lines connect the BB parameters corresponding to four different levels of $L_{\rm X}$ (in erg s$^{-1}$). \textit{Right}: X--ray luminosity (in the 2--10 keV energy range) of the pulsars with a detected thermal excess as a function of the pulse period; \textit{filled circles} refer to the \textit{hot--BB} pulsars, \textit{empty squares} to the \textit{soft--excess} sources.
}
\label{figure}
\end{figure*}

The \XMM\ spectra of these Be pulsars cannot be described with a single-component model: the fits with a power--law (PL) or a blackbody (BB) model are affected by large residuals, while other models are rejected by the data. On the other hand, a good fit is obtained with a PL+BB model. In all cases the BB component is characterized by a high temperature ($kT_{\rm BB} > 1$ keV) and a small emission radius ($R_{\rm BB} < 0.5$ km), and contributes to 30--40 \% of the source flux below 10 keV (Table~\ref{parameters}): therefore this \textit{hot--BB} spectral component can be considered as an additional common property of the persistent Be pulsars, which stands beside those previously known.

\begin{table*}
\caption{Main parameters of the persistent Be X--ray binaries}\label{parameters}
\vspace*{-0.75 cm}
\begin{center}
\begin{tabular}{lcccc}
\hline
Source				& 4U 0352+309		& RX J0146.9+6121	& RX J1037.5-5647	& RX J0440.9+4431	\\
\hline
Optical Spectral Type		& 09.5 IIIe		& B0 IIIe		& B0 III-Ve		& B0.2 Ve	\\
Distance (kpc)			& $\simeq$ 1		& $\simeq$ 2.5		& $\simeq$ 5		& $\simeq$ 3.3	\\
Pulse Period (s)		& 839.3			& 1396.1		& 853.4			& 204.98	\\
$L_{\rm X}$ (0.3--10 keV, erg s$^{-1}$)			& $1.4\times10^{35}$	& $1.5\times10^{34}$	& $1.2\times10^{34}$	& $8.3\times10^{34}$	\\
$L_{\rm BB}$ (0.3--10 keV, erg s$^{-1}$)		& $5.5\times10^{34}$	& $3.6\times10^{33}$	& $5.0\times10^{33}$	& $2.9\times10^{34}$	\\
$L_{\rm BB}/L_{\rm X}$ (\%)	& 39			& 24			& 42			& 35			\\
$T_{\rm BB}$ (keV)		& 1.42 $\pm$ 0.03	& 1.11 $\pm$ 0.06	& 1.26$^{+0.16}_{-0.09}$& 1.34 $\pm$ 0.04	\\
$R_{\rm BB}$ (m)		& 361 $\pm$ 3		& 140 $\pm$ 15		& 128$^{+13}_{-21}$	& 273 $\pm$ 16		\\
$R_{\rm col}$ (m)		& $\sim$ 330		& $\sim$ 230		& $\sim$ 200		& $\sim$ 320		\\
\hline
\end{tabular}
\end{center}
\vspace*{-0.5 cm}
\end{table*}

Based on the emission models proposed by \citet{Hickox+04}, the low luminosity of these pulsars suggests that the observed \textit{BB} component is due to thermal emission from the NS polar cap. Assuming the standard NS parameters $M_{\rm NS}$ = 1.4 $M_{\odot}$, $R_{\rm NS}$ = $10^6$ cm, and $B_{\rm NS} = 10^{12}$ G, from the source luminosities we can estimate the accretion rate and, then, the radius of the accretion column $R_{\rm col}$. For all sources we found that $R_{\rm col} \sim R_{\rm BB}$, thus confirming the previous hypothesis.

A spectral feature similar to the \textit{hot BB} of the persistent Be pulsars has been observed also in other low--luminosity ($L_{\rm X} \le 10^{36}$ erg s$^{-1}$) high--mass X--ray binaries \citep{LaPalombara+12}. In Fig.~\ref{figure} (\textit{left panel}) we report the best--fit radius and temperature for the BB component of these sources: it shows that, for all the sources, $kT_{\rm BB} >$ 0.5 keV and $R_{\rm BB} <$ 2 km; moreover, for the less luminosity sources ($L_{\rm X} \sim 10^{34}$ erg s$^{-1}$), the spectral parameters are within a narrow range of values ($kT_{\rm BB} \sim$ 1--2 keV and $R_{\rm BB} <$ 200 m), with a 20--40 \% contribution of the blackbody component.

In contrast to this sample of sources, several pulsars are characterized by a \textit{soft} excess, since the fit of this component with a thermal emission model provides low temperatures ($kT_{\rm SE} < 0.5$ keV) and large emitting regions ($R_{\rm SE} > 100$ km). In Fig.~\ref{figure} (\textit{right panel}) we report the luminosity and pulse period of both types of pulsars. On the average, the \textit{hot--BB} pulsars are characterized by the lowest luminosities and the longest periods; their \textit{hot--BB} spectral component is a common feature which separates them from all the other pulsars, strongly suggesting that they form a distinct and well--defined class of binary pulsars.
\vspace*{-0.75 cm}

\bibliographystyle{aa}

\end{document}